\documentclass[12pt]{article}
\newif\iffigs\figstrue
\usepackage{latexsym}
\iffigs
  \input{epsf}
\else
\fi
\font\tenmsbm=msbm10 scaled 1200
\font\sevenmsbm=msbm9
\newfam\msbmfam
\textfont\msbmfam=\tenmsbm
\scriptfont\msbmfam=\sevenmsbm
\def\msbm{\fam\msbmfam\tenmsbm}

\makeatletter
\@addtoreset{equation}{section}
\makeatother

\newsavebox{\uuunit}
\sbox{\uuunit}
    {\setlength{\unitlength}{0.825em}
     \begin{picture}(0.6,0.7)
        \thinlines
        \put(0,0){\line(1,0){0.5}}
        \put(0.15,0){\line(0,1){0.7}}
        \put(0.35,0){\line(0,1){0.8}}
       \multiput(0.3,0.8)(-0.04,-0.02){10}{\rule{0.5pt}{0.5pt}}
     \end {picture}}

\def\IP{\relax{\rm I\kern-.18em P}}

\def\inbar{\vrule height1.5ex width.4pt depth0pt}
\def\IC{\relax\,\hbox{$\inbar\kern-.3em{\rm C}$}}
\def\bfzero{\relax\,\hbox{$\inbar\kern-.3em{\rm 0}$}}

\def\bfone{\relax{\rm 1\kern-.35em 1}}

\def\tilde{\widetilde}

\def\IE{\relax{{\rm I\kern-.18em E}}}

\def\IGam{\relax{{\rm I}\kern-.18em \Gamma}}

\def\bet{\begin{tabular}}
\def\eet{\end{tabular}}

\def\uA{\underline{A}}

\def\ha{\hat{a}}
\def\hb{\hat{b}}

\begin{document}
\begin{titlepage}
\begin{flushright}
Preprint DFTT 99/12 \\
hep-th/9903041\\
\end{flushright}
\vskip 2cm
\begin{center}
{\Large \bf  The 
singleton action\\
\vskip 1.5mm
from the supermembrane\footnote{Talk 
given by P. T. at the TMR meeting 
of the project 
``{\sl Quantum Aspects of Gauge Theories, Supersymmetry and Unification}''
 TMR contract
 ERBFMRX-CT96-0045, Corfu September 20--26, 1998. To appear in the proceedings. 
\\
This  work is also supported in part by EEC  under TMR contract
 ERBFMRX-CT96-0012, in which M. Trigiante is associated to Swansea
University.} }\\

\vfill
{  Gianguido Dall'Agata$^2$, Davide Fabbri$^2$,
   Christophe Fraser$^3$, \\
   Pietro Fr\'e $^2$, Piet Termonia$^2$
   and  Mario Trigiante$^4$   } \\
\vfill
{\small
$^2$ Dipartimento di Fisica Teorica, Universit\'a di Torino, via P. 
Giuria 1,
I-10125 Torino, \\
 Istituto Nazionale di Fisica Nucleare (INFN) - Sezione di Torino, 
Italy \\
\vspace{6pt}
$^3$  Wellcome Trust Centre for the Epidemiology of Infectious Diseases,
    University of Oxford,
    South Parks Road,
    OXFORD OX1 3PS, UK.\\
\vspace{6pt}
$^4$ Department of Physics, University of Wales Swansea, Singleton
Park,\\
 Swansea SA2 8PP, United Kingdom\\
}
\end{center}
\vfill
\begin{abstract}
We derive the free $Osp(8|4)$ singleton action by sending the $M2$ brane 
to the Minkowski boundary of an $AdS_4 \times {\cal M}_7$ background.
We do this by means of the solvable Lie algebra parametrization
of the coset space. We also give some comments on 
singleton actions from membranes on $AdS_4 \times G/H$ backgrounds.
\end{abstract}
\vspace{2mm} \vfill 
\end{titlepage}
\section{Introduction}
For about a year now, 
there has been a revival of interest in supergravity vacua of the form
\begin{eqnarray}
AdS \times {\cal M}\,,
\end{eqnarray}
where $AdS$ is an 
anti--de Sitter space and ${\cal M}$ a compact Einstein manifold.
This revival started after Maldacena's conjecture \cite{Maldacena} 
of duality between Kaluza--Klein supergravity theories 
in the bulk of anti--de Sitter space and conformal field theories on the boundary,
\begin{equation}
\label{holography}
\mbox{KK on $AdS$} \,\, / \,\, \mbox{CFT on $\partial AdS$} \,. 
\end{equation}
Since the proposal by Gubser, Klebanov, Polyakov and Witten 
 for this duality \cite{GKPW}, some work is 
being done with the scope of testing this conjecture of holography.
This has brought the whole research area of 
branes and AdS representations from the eighties back alive.

Of particular interest in testing this AdS/CFT duality is the singleton problem.
Representations of AdS algebras 
have been studied extensively in the past \cite{AdSers}.
They are characterized by a lowest energy  and the total angular momentum. 
Most of them have a Poincar\'e analogue, except for some ultra--short
representations which are referred to as singleton representations. The 
most considerable property of these singletons is that they can not be 
formulated as a field theory on the bulk of the AdS space. Yet, singleton 
actions exist, but can only be formulated if the singleton fields are 
restricted to live on the boundary of $AdS$. 
It has also been known for a 
long time how to get these singleton actions and what they describe: 
namely, they describe the small fluctuations of a brane at {\it the end of the universe} \cite{fieldreal}. All this has been known for about ten years. Still, 
the singleton has not been constructed explicitly from the brane until
recently. It is clear that an explicit realization of this singleton
at the end of the universe is an important ingredient for testing the 
AdS/CFT conjecture.

Thus, to find the singleton from the super membrane one has to do the 
following:
\begin{enumerate}
\item{Consider the super membrane action that is invariant with respect to $\kappa$ supersymmetry.}
\item{Expand this action around a classical solution.}
\item{Send it to infinity.}
\end{enumerate}

As far as the first step is concerned,
membrane actions in an explicit background of 
$AdS_4\times S^7$ have  been constructed by our collaboration \cite{torinos7}
and  others \cite{braness7}.
Here, we will overview the construction of the membrane action and
the derivation the singleton action in this background,
as it was done in \cite{torinos7}.
However, 
before carrying out this programme let us point out that
the singleton theory that is retrieved on this space is quite trivial
and doesn't yield a proper test  for the AdS/CFT correspondence.
Therefore one has to consider the singleton problem on 
other non--trivial backgrounds. 
Suitable backgrounds for this are given 
if one replaces the sphere by other 
coset manifolds $G/H$. A complete classification
of these backgrounds for $D=11$ is already know and these spaces have
been thoroughly studied in the eighties \cite{G/H}. 
They are in one--to--one 
correspondence with the $G/H$ Freund--Rubin compactifications.
Also, the number of preserved supersymmetries $N$ are known. 

Of all these coset spaces, 
the case of the round and squashed seven spheres are the best known 
(corresponding to $N=8$ \cite{dewit2} and
$N=1$ near horizon supergravities) 
but in the eighties the Kaluza--Klein spectra have been 
systematically
derived  also for all the other solutions using the technique of 
harmonic expansions
\cite{mpqr1}. The organization of these spectra in 
supermultiplets
is known not only for the round $S^7$ \cite{gunaydin} but also for
the case of supersymmetric $M^{pqr}$ spaces
$$
 M^{pqr} \equiv  \frac{SU(3) \,\times \, SU(2) \, \times \, 
U(1)}{SU(2) \, \times \,
 U(1) \, \times \, U(1)}
\,,
$$
where $p,q,r\in \hbox{\msbm Z}$ define the embedding of the
$U(1)^2$ factor of $H$ in $G$.  For $p=q=\mbox{odd}$ we have 
$N=2$, in all the other (non supersymmetric) cases
we have $N=0$.
The $N=2$ multiplet structure was obtained
in \cite{mpqr4}. At present a group in Torino \cite{torinogroup}
is doing the harmonic analysis on the so--called Stiefel and $N^{010}$
manifolds as well.

Since much is known and will be known about these spaces, they
constitute an excellent laboratory to make direct checks of the
holographic correspondence.

Let's now clarify the qualitative difference between 
the seven sphere and the other $G/H$ spaces. For a $G/H$ space that admits 
$N$ supersymmetries, the isometry group is factorized as follows:
$$
G \, = \, G^\prime \, \otimes \, SO\left( N \right) \,,
$$
where $SO\left( N \right)$ is the $R$--symmetry 
of the orthosymplectic algebra $Osp \left( N 
\vert 4
\right )$, while the factor $G^\prime$ is the gauge--group of the
vector multiplets. Correspondingly the three--dimensional 
world--volume action
of the $CFT$ must have the following superconformal symmetry:
\begin{equation}
\begin{array}{ccc}
Osp \left( N \vert 4 \right) \, \times
\,
G^\prime \\
\end{array} \,,
\label{scGH}
\end{equation}
where $G^\prime$ is a {\it flavor group}.
In the maximal case the harmonics on $S^7$
are labeled only by $R$--symmetry representations while
in the lower susy case they  depend both on
$R$ labels and on representations of the gauge/flavour group 
$G^\prime$.
The structure of $Osp(8\vert 4)$ supermultiplets
determines completely their $R$--symmetry representation content so 
that the harmonic
analysis becomes superfluous in this case. The eigenvalues of the 
internal laplacians
which determine {\it the Kaluza--Klein masses} of the $Osp(8\vert 4)$ 
graviton
multiplets or, in the conformal reinterpretation of the theory, the 
{\it conformal
weights} of the corresponding primary operators, are already fixed by 
supersymmetry
and need not be calculated. In this sense the
correspondence (\ref{holography}) is somewhat trivial in the maximal
susy case: once the superconformal algebra has been
identified with the super-isometry group $Osp(8\vert 4)$ the
correspondence between conformal weights and Kaluza--Klein masses is
simply guaranteed by representation theory of the superalgebra.
On the other hand in the lower susy case the structure of  the
$Osp(N\vert 4)$ supermultiplets fixes only their content in
$SO \left( N \right)$ representations while the Kaluza--Klein
masses, calculated through harmonic analysis depend also on
$G^\prime$ labels. In this case the holographic correspondence yields
a definite prediction on the conformal weights that, as far as 
superconformal symmetry
is concerned would be arbitrary. Explicit verification of these
predictions would provide a much more stringent proof of the
holographic correspondence and yield a deeper insight in its inner
working. However in order to set up such a direct verification one
has to solve a problem that was left open in Kaluza--Klein
supergravity: the singleton problem.

In the remainder of this text we will restrict ourselves to
the membrane on the seven sphere. The case of the cosets 
with lower supersymmetry is currently under investigation.

To avoid any confusion, we would like to stress here that we call 
singleton field theory the flat space limit of the free field theory  
of \cite{QFT,Nicolai}.
We point out that, since we are going to find a theory living on a 
three--dimensional Minkowski space rather than on $S^2 \times S^1$,
we have no scalar mass term which 
was instead required in \cite{QFT,Nicolai} for conformal invariance.
We will see that it can also be derived as the theory living
on the solitonic M2 brane. 
 
\section{The supermembrane on $AdS_4 \times S^7$}
We consider the space $AdS_4 \times S^7$, with given metric,
\begin{equation}
\label{solvmet}
d\tilde{s}^2 =  \rho^2 \left( -dt^2 + dx^2 + dw^2 \right) + 
\frac{1}{\rho^2} d\rho^2 + 4 d\Omega_7^2\,,
\end{equation}
with coordinates, 
$$
\cases{ \rho  \, \in \ ]0,\infty [\cr
t,w,x \, \in \ ]\!-\!\infty , \infty [ }
\,,
$$
and $d\Omega_7^2$ is the metric of the seven sphere. 
This is the near horizon geometry of the $M2$ brane \cite{nearhorizongeom}. 
Moreover in \cite{stablequantum} it was shown that this
is a stable quantum vacuum of of the $11D$ supergravity.
The $AdS$ superspace is defined as the following coset
$$
AdS^{(8|4)} = \frac{OSp(8|4)}{SO(1,3) \otimes SO(8)}
\label{coset}
$$
and it is spanned by the four coordinates of the $AdS_4$ manifold
and by eight Majorana spinors (i.e. they have 32 real components)
parametrizing the fermionic generators 
 of the superalgebra.

This space can be described by means of a super solvable Lie 
algebra parametrization. To see what this solvable Lie 
algebra parametrization is, let's have a look at
the familiar ``Union Jack'' root diagram of $C_2$, 
which is the complexification of $SO(2,3)$ shown in figure 1. 
The fermionic supercharges form
a square weight diagram within this figure, and the supertranslation 
algebra
is then simply that the anticommutator of two fermions is given by 
vector
addition of the corresponding weights in the diagram. The diagram can 
in fact
be seen as a projection of the full $Osp(8|4)$ root diagram, since 
the $SO(8)$
roots lie on a perpendicular hyperplane, and so on this diagram they 
would be at the
centre.

%
\begin{figure}
\epsfxsize=8cm \epsfysize=2.0in  
\centerline{\epsfbox{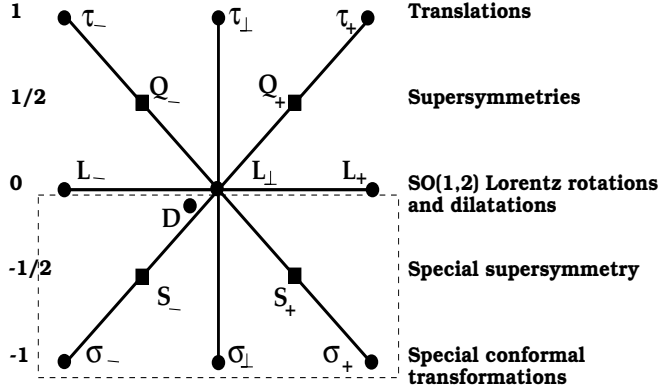}}
\caption{ {\small The root diagram of $SO(2,3)$. The bosonic weights 
are represented
by circles, and the fermionic weights by squares. The dilatation 
charge of
horizontal planes in the diagram are on the left, while the 
worldvolume
theory interpretation of the planes of generators are labelled to the 
right.
The supersolvable algebra is the boxed subalgebra. }}
\unitlength=1mm
\end{figure}
It is now easy to see that the generators in the box 
$\{ S_{\pm}, \sigma_{\pm}, \sigma_{\perp}, D\}$, form a super solvable Lie 
algebra with non--zero (anti) commutation relations,
\begin{equation}
\{S,S\}\sim \sigma\,, \qquad [D,S]\sim S\,, \qquad [D,\sigma] \sim \sigma\,,
\label{solvecomm}
\end{equation}
since its second derivative is zero. The coset representative is now obtained 
by exponentiating the product of these generators with the coordinates 
of the coset space. We choose to write the coset representative
as $L=L_F L_B$ with
\begin{eqnarray}
L_B &=& \exp(\rho D) \, \exp(\sqrt{2}x{\cal \sigma}_{\perp}+
t({\cal \sigma}_{+}+{\cal \sigma}_{-})+
w({\cal \sigma}_{+}-{\cal \sigma}_{-})) \,,\nonumber \\
L_F &=&
\exp\left( \theta_1^A S_1^A + \theta_2^A  S_2^A \right)\,,
\label{exponentiation}
\end{eqnarray}
where $\rho, x, t, w, \theta^A$ are the bosonic and fermionic coordinates.
From the left invariant form,
\begin{eqnarray}
\label{liform}
\Omega &=& L^{-1}dL=\Omega_B + L_B^{-1} \Omega_F L_B\,,
\nonumber \\
\Omega_F &=& L_F^{-1}d L_F\,,
\end{eqnarray}
one derives the vielbeins\footnote{
For the fermionic coordinates the following notation is understood:
 $\bar\theta^A = \theta^A \gamma^0$ and
\begin{equation}
\theta^A = 
\; \left( \begin{tabular}{c}  $\theta_1^A$ \\
$ \theta_2^A$ \\ $0$ \\ $0$ \end{tabular} \right)\,.
\end{equation}
   },
\begin{eqnarray}
E^0 &=& - \rho dt -  \rho \bar\theta^A \gamma^0 d \theta^A, \nonumber \\
E^1 &=& \rho dw -  \rho \bar\theta^A \gamma^1 d \theta^A, \nonumber \\
E^2 &=&  \frac{1}{2 \rho} d\rho,  \label{param1} \\
E^3 &=& \rho dx -  \rho \bar\theta^A \gamma^3 d \theta^A, \nonumber
\end{eqnarray}
and
\begin{equation}
\psi^A = \sqrt{2e\rho}
\; \left( \begin{tabular}{c} $0$ \\ $0$ \\ $d\theta_1^A$ \\
$d \theta_2^A$ \end{tabular} \right),
\end{equation}
Notice that, due to the (anti) commutation relations (\ref{solvecomm}) of the 
solvable Lie algebra parametrization, the exponentiation (\ref{exponentiation})
only contains a finite number of terms. Hence, it immediately follows that 
the vielbeins are at most quadratic in their anti--commuting 
coordinates. For a discussion on this see \cite{torinos7, secorder}.
 
Another convenient feature of the solvable Lie algebra parametrization is that
one projects out half of the spinors. This is equivalent to the projection
of the $\kappa$--symmetry operator 
and thus at this stage the $\kappa$ symmetry has already been fixed.

This parametrization of the manifold gives rise to the metric (\ref{solvmet}).

To complete the parametrization of the target superspace one still has to 
define the vielbeins of the seven sphere.
To do so, we
call $y^{\hat a}$ the seven coordinates of the sphere and write 
in  stereographic projection coordinates:
\begin{equation}
\label{param5}
E^{\hat a} = 
- \delta ^{\hat a}_{\ \hat m} \frac{d y^{\hat m}}{1 + y^2}.
\end{equation}

Beside the $\kappa$ symmetry we also fix 
the three--dimensional world--volume diffeomorphisms 
imposing the static gauge choice.
To obtain static solutions we have to identify 
\begin{equation}
\xi^I \equiv (-t,w, x),
\label{orfan30}
\end{equation}
where $I=0,1,2$, is the curved index of the brane.

For the action of the membrane we get:
\begin{equation}
\label{confaction}
S = 2 \int \sqrt{-\mbox{det}(h_{IJ})} \, d^3\xi + 4!4! \int B\,,
\end{equation}
with\footnote{
where $i=0, 1, 2$, $\epsilon_{012}=1$ and $\eta_{IJ}=diag\{-++\}$. 
The  seven--dimensional gamma matrices $\tau^{\hat a}$ are
the generators of the $SO(7)$ Clifford algebra and
$\eta_A$ are the eight real killing spinors on the seven sphere.
For details on the conventions see \cite{torinos7}.
}
\begin{eqnarray}
h_{IJ}(\xi) &=& 
\frac{1}{4 \rho^2}
\partial_I \rho \partial_J \rho + \frac{1}{(1+y^2)^2} 
\partial_I 
y^{\ha} \partial_J y_{\ha}  \nonumber \\
\label{metrica}
&& + \rho^2 \left(\eta_{IJ} - 2 e \bar\theta^A \gamma^i \partial_{(J} \theta^A 
\delta_{i\,I)} +  \bar\theta^A \gamma^i \partial_I \theta^A \ \bar\theta^B \gamma_i
\partial_J \theta^B \right).
\end{eqnarray}
The expression for $B$ is given by
\begin{eqnarray}
B &=& \frac{d^3 \xi}{4!4!}\left[ \epsilon^{IJK} \rho^3 (\delta_I{}^i - 
 \bar\theta^A \gamma^i \partial _I \theta^A)
(\delta_J{}^j -  \bar\theta^A \gamma^j \partial_J \theta^A)
(\delta_K{}^k -  \bar\theta^A \gamma^k \partial _K \theta^A)
\frac{\epsilon_{ijk}}{3} + \right. \nonumber \\
&& \hskip .9cm 
- \left. \frac{1}{(1+y^2)} \epsilon^{IJK} \partial _I y^{\ha} \;  
\eta_A \tau_{\ha} \eta_B
\; \rho \partial _J \bar\theta^A \partial _K \theta^B \right] .
\label{orfan35}
\end{eqnarray}

The isometries of (\ref{coset})
can be calculated explicitly in this parametrization.
As was noted in \cite{Maldacena, isos}, they realize conformal symmetry
on the brane.
For example, for the dilatation one finds, 
\begin{eqnarray}
\delta \rho &=& \rho, \label{deltarho}\\
\delta \xi^I &=&  - \xi^I, \label{dilat} \\
\delta \theta^A_{\alpha} &=& - \frac{1}{2}\theta^A_{\alpha}\,.
\end{eqnarray}

\section{The singleton action from the supermembrane}

In order to retrieve the $Osp(8|4)$ singleton action we now have to do the following.
First we consider a classical solution of the action (\ref{confaction}),
\begin{equation}
 \xi^I \equiv (-t,w,x), \quad \partial _I y^{\ha} = 0,
\quad \theta^A = 0, \quad \rho=\bar\rho={\rm const} \,,
\label{orfan39}
\end{equation}
then expand the transverse coordinates to the brane
around the values for this classical solution. For this we use normal coordinates.
Thus we write
\begin{eqnarray}
\rho &=& \bar{\rho} + {\alpha'}^{\frac{3}{2}} \tilde\rho\,, 
\nonumber\\
\label{fluttua}
y^{\ha} &=& {\alpha'}^{\frac{3}{2}} \tilde y^{\ha} \,, \\
\theta^A &=& {\alpha'}^{\frac{3}{2}} \Theta^A\,, \nonumber 
\end{eqnarray}
where $\tilde\rho, \tilde y^{\ha}, \Theta^A$ represent the fluctuations
and $\alpha'$ is the membrane tension.
Thus the action (\ref{confaction}) gets the following expansion
as a power series in $\alpha^\prime$:
\begin{equation}
\label{orfan41}
{\cal L} = \sum_{n = 0}^{\infty} \alpha^{\prime}{}^{\frac{3(n-2)}{2}}
{\cal L}_{(n)} \,,
\end{equation}
with
\begin{equation}
\label{orfan42}
{\cal L}_{(0)} = 0 = {\cal L}_{(1)} \,,
\end{equation}
and we are to recover the singleton action from the order 1 term
\begin{eqnarray}
{\cal L}_{(2)} &=& \frac{1}{4 \bar{\rho}} \eta^{IJ} 
\partial_I \tilde\rho \,
\partial_J \tilde\rho + \bar{\rho} \eta^{IJ} 
\partial_I \tilde y^{\ha} \,
\partial_J \tilde y^{\hb} \delta_{\ha\hb} - 
2 \, \bar{\rho}^3 \; \bar{\Theta}^A \hat\sigma^i
\partial_I \Theta^A \delta_i^I.
\label{orfan47}
\end{eqnarray}
The final step  is to send the brane to the boundary of AdS. 
The boundary of AdS lies at
\begin{equation}
\bar\rho \to \infty \quad \hbox{ and } \quad \bar\rho \to 0\,,
\end{equation}
which is a conformally compactified
Minkowski space (see one of the appendices in \cite{torinos7}). 
It is already clear from the form of the action (\ref{orfan47}) that 
in order to take one of these limits one has to rescale the fields
$\tilde\rho, \tilde y^{\ha}$ and $\Theta^A$. In fact, a proper analysis 
of the supersymmetry variation, which we do not present here but can be found in
\cite{torinos7}, shows us that these rescalings have to be done 
according to
\begin{equation}
\lambda = \bar{\rho}^{\frac{3}{2}} \Theta^A, \quad \tilde{P} = 
\frac{\tilde{\rho}}{\sqrt{\bar{\rho}}}, \quad \tilde{Y}^{\ha} = 
\sqrt{\bar{\rho}} \tilde{y}^{\ha}.
\end{equation}
Using the notation
\begin{equation}
Y^{\uA} \equiv \left\{\frac{\tilde{P}}{4}, \frac{\tilde{Y}}{2} 
\right\} \,,
\end{equation}
the action (\ref{orfan47}) becomes
\begin{equation}
\label{singleton}
{\cal L} =  4 \, \eta^{IJ} \, \partial_I Y^{\uA} \partial_J 
Y^{\uA} 
- 2 \, \bar\lambda^A \hat\sigma^I \partial_I \lambda^A \,.
\end{equation}
Clearly, this action  has the right form to become the singleton action.
Yet, for generic values of $\bar\rho$ it does not. To see this, let's look at the
dilatation symmetry of the action (\ref{confaction}).
The transformation (\ref{deltarho}) can only become a symmetry of the
action (\ref{confaction})
if we place the brane at the boundary. 

So we conclude that the singleton is found after putting the brane at the boundary
of the Anti--de Sitter space and that
the singleton field theory describes the centre of mass
degrees of freedom of the $M2$ brane.
\vskip 1cm
{\bf Acknowledgement}
P. T. is grateful  P. Claus and D. Sorokin for useful discussions during the 
conference.
%
%

\end{document}